\documentclass[12pt]{article}
\usepackage{amsmath}
\usepackage{cite}
\usepackage{epsfig}
\usepackage{epstopdf}
\usepackage{amssymb}

\def\[{\left [}
\def\]{\right ]}
\def\({\left (}
\def\){\right )}

\def\r{\rho}

\def\r2{\sqrt{2}}

\newcommand{\bbibitem}[1]{\bibitem{#1}\marginpar{#1}}

\def\Label#1{\label{#1}%
  \smash{\hbox to0pt{\raise1ex\hbox{\tiny[#1]}\hss}}}
\def\noLabels{\let\Label=\label}
\def\nobbibitem{\let\bbibitem=\bibitem}

\newcommand{\bea}{\begin{eqnarray}}
\newcommand{\eea}{\end{eqnarray}}

\newcommand{\beq} {\begin{equation}}
\newcommand{\eeq} {\end{equation}}
\newcommand{\beqa} {\begin{eqnarray}}
\newcommand{\eeqa} {\end{eqnarray}}

\newcommand{\beqn}{\begin{eqnarray}}
\newcommand{\eeqn}{\end{eqnarray}}

\begin{document}

\begin{flushright}
Nordita-2012-92\\
RH-11-2012\\
\end{flushright}

\vskip 2cm \centerline{\Large {\bf Non-equilibrium Wilson loops in $\mathcal{N}=4$ SYM}}
\vskip 2cm
\renewcommand{\thefootnote}{\fnsymbol{footnote}}
\centerline
{{\bf Ville Ker\"{a}nen,$^{1,2}$
\footnote{vkeranen@nordita.org}
}}
\vskip .5cm
\centerline{\it
${}^{1}$Nordita, KTH Royal Institute of Technology and Stockholm University}
\centerline{\it Roslagstullsbacken 23, SE-106 91 Stockholm, Sweden}
\centerline{\it
${}^{2}$ University of Iceland, Science Institute}
\centerline{\it Dunhaga 3, IS-107 Reykjavik, Iceland}

\setcounter{footnote}{0}
\renewcommand{\thefootnote}{\arabic{footnote}}

\begin{abstract}
We consider rectangular Wilson loops in certain non-equilibrium quantum states in $\mathcal{N}=4$ SYM at weak coupling, prepared with a quantum quench. We find that in the ladder approximation, the Bethe-Salpeter equation can be reduced to solving a massive 1+1 dimensional wave-equation with a leaking boundary condition leading to a quasinormal behavior analogous to what is found in studying dynamics of fields in black hole backrounds. Furthermore, we find that the Wilson loops with size $L$ approach a thermal form after time $T\approx L/2$. The thermal form found in the current paper follows from the particular initial state chosen.
\end{abstract}
\newpage
\tableofcontents

\section{Introduction}

There is a large amount of evidence that $\mathcal{N}=4$ super Yang Mills theory is secretly a string theory in $AdS\times S^5$, a quantum theory of gravity. One of the main differences between gravitational theories and non-gravitational
theories is the amazing property that in gravitational theories the causal structure of the spacetime is dynamical. A manifestation of this difference is black holes, which have the property of absorbing field fluctuations never letting them back to the outside spacetime\footnote{This is when considering large black holes in $AdS$ which do not evaporate.}.

Static black hole geometries are dual to thermal states in the field theory. A qualitative understanding of many of the features of
black holes can been understood in terms of large $N$ gauge theory in a thermal state. Black hole formation in the bulk is expected to be dual to
thermalization of an excited state in strongly coupled SYM. Motivated by black hole physics,
non-equilibrium dynamics of large-$N$ matrix theories has become an interesting problem, studied
for example in \cite{Festuccia:2006sa,Asplund:2008xd,Iizuka:2008hg,Asplund:2011qj,Asplund:2011cq}.
The purpose of this work is to initiate a study of non-equilibrium dynamics
of Wilson loops in weakly coupled $\mathcal{N}=4$ SYM. Our aim here is to study whether the weak coupling
dynamics has any qualitative resemblance with expectations from gravitational physics.

We consider deforming SYM with a time dependent source for the Konishi (mass) operator Tr$\Phi^I\Phi^I$. More specifically, we start with a non-vanishing value of the source and suddenly turn it off. This way the subsequent time evolution is generated by the usual SYM Hamiltonian. The state is a high
energy state which can be expected to thermalize. This procedure is an example of a quantum quench. The behavior of correlation functions and entanglement entropy in conformal (and other) field theories subject to a quantum quench have been studied for example in \cite{Calabrese:2005in,Calabrese:2007rg,Sotiriadis:2010si}. In this work we extend
some of these methods to loop operators in weakly coupled matrix field theory. In particular, these loop operators can be obtained as a limit of the Maldacena-Wilson loop in $\mathcal{N}=4$ SYM. In what follows we will use the terminology that thermalization of an observable means that it reaches a value identical to the corresponding value in a thermal state.
Thus, we do not make a distinction between different ways of reaching the thermal form, e.g. by interactions of particles or by choice of the initial state.

To calculate the Wilson loop we use two approximations. The first one was introduced in \cite{Correa:2012nk} to isolate the scalar ladder Feynman diagram contributions to the Wilson loop. We will call this the ladder approximation. The idea is to analytically continue the Maldacena-Wilson loop into a case where the $SO(6)$ orientations of the external
fundamental charges are separated by a large imaginary angle. An alternative way of interpreting the current paper is as a study of loop operators in large-$N$ matrix field theory of
free scalar fields only. In this free theory, the ladder diagrams are the only ones present at large-$N$. \footnote{Somewhat surprisingly the results from the ladder approximation agree with strong coupling results in the vacuum (when the coupling constant is properly defined through analytic continuation) and an impressive agreement with string theory results is found \cite{Erickson:1999qv,Erickson:2000af,Semenoff:2002kk,Correa:2012nk}. Also first order corrections to the leading ladder approximation are found to agree with the strong coupling results \cite{Bykov:2012sc,Henn:2012qz}. It should be emphasized that there is no clear reason for the agreement between the ladder approximation and bulk string theory to continue to
other states than the vacuum state.}

As a second approximation we take a limit in which the initial value of the scalar mass (the source of the Konishi operator) is large as compared to the inverse separation $1/L$ of the timelike segments in the Wilson loop. This approximation was introduced in \cite{Calabrese:2007rg,Sotiriadis:2010si} to study scalar two point correlations in such a non equilibrium state. Following the terminology of \cite{Calabrese:2007rg,Sotiriadis:2010si} we will call this the deep quench approximation. In this approximation the two point function of the scalar field simplifies drastically allowing for analytic calculations.

By summing up the ladder diagrams we find that the Bethe-Salpeter equation can be written as a 1+1 dimension Klein-Gordon equation with a spacetime dependent potential
determined by the free field scalar two point function. We study the solutions to the Klein-Gordon equation both using analytic and numeric tools.
The results are discussed in the discussion section.

\section{Setting up the non-equilibrium state}

We start from $\mathcal{N}=4$ SYM with a time dependent source turned on for the Konishi operator Tr$(\Phi^I\Phi^I)$,
\beq
S=-\frac{1}{g^2}\int d^4x\textrm{Tr}\Big(\frac{1}{2}F^2+(D\Phi^I)^2+m^2(t)(\Phi^I)^2-\[\Phi^I,\Phi^J\]^2+\textrm{fermions}\Big).
\eeq
A convenient choice for calculations is to choose $m^2(t)=\theta(-t)m_0^2$. We believe that a more generic choice of $m(t)$ interpolating between $m_0$ and $0$ will lead to qualitatively similar results.

We will begin from the ground state at times $t<0$. In the $g\rightarrow 0$ limit this state is annihilated by the corresponding
annihilation operators of the free SYM fields. In momentum space, and in the Heisenberg picture, the free equations of motion for $\Phi$ at times $t<0$ are solved by
\beq
\Phi(t,k)=\frac{g}{\sqrt{2\omega_0(k)}}\Big(Be^{-i\omega_0(k)t}+B^{\dagger}e^{i\omega_0(k) t}\Big).
\eeq
After $t>0$, the corresponding solution is
\beq
\Phi(t,k)=\frac{g}{\sqrt{2\omega(k)}}\Big(Ae^{-i\omega(k)t}+A^{\dagger}e^{i\omega(k) t}\Big).
\eeq
The factors of the coupling constant $g$ are chosen for normalization. Above we have introduced the notation $\omega_0(k)=\sqrt{k^2+m_0^2}$ and $\omega(k)=|k|$. Matching these two gives us a Bogoliubov transformation relating the two sets of creation and annihilation operators
\beq
A=\alpha B+\beta B^{\dagger},\quad A^{\dagger}=\alpha^* B^{\dagger}+\beta^* B\label{eq:bog}
\eeq
where
\beq
\alpha=\frac{1}{2}\Big(\sqrt{\frac{\omega}{\omega_0}}+\sqrt{\frac{\omega_0}{\omega}}\Big),\quad \beta=\frac{1}{2}\Big(\sqrt{\frac{\omega}{\omega_0}}-\sqrt{\frac{\omega_0}{\omega}}\Big).
\eeq
Thus, the quench induces a non-vanishing particle number
\beq
\langle N_k\rangle=|\beta|^2=\frac{(k-\sqrt{k^2+m_0^2})^2}{4k\sqrt{k^2+m_0^2}}.
\eeq
This can be compared to a thermal ensemble $\langle N_k\rangle=(e^{\beta k}-1)^{-1}$. At small momenta these have the same form
$\langle N_k\rangle\approx 1/(k\beta)$, if we would identify $m_0/4$ as an effective temperature. The particle number differs significantly from the
thermal one at large momenta as $\langle N_k\rangle\approx m_0^4/(16 k^4)$, which in the thermal case would vanish as $e^{-\beta k}$.

\begin{figure}[h]
\begin{center}
\includegraphics[scale=.5]{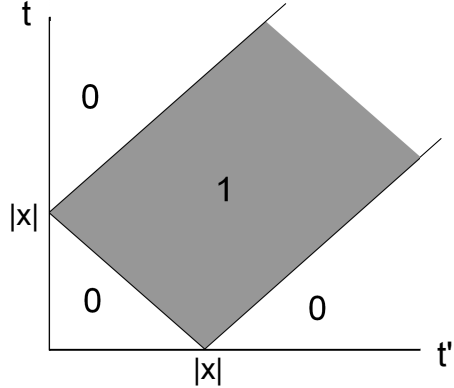}
\caption{\label{fig:step} The function $\Theta(|x|,t,t')$ visualized on $t,t'$ plane for a fixed value of $|x|$.}
\end{center}
\end{figure}

The two point function in this time dependent state can be worked out by using the fact that the quench state satisfies $B|\psi\rangle=0$, which gives
\begin{align}
&\langle \Phi_{a}^{I}(t,x)\Phi_{b}^{J}(t',0)\rangle=g^2\delta_{ab}\delta^{IJ}\times,\nonumber
\\
&\times\int\frac{d^3k}{(2\pi)^3}\frac{e^{i k\cdot x}}{2|k|}\Big(|\alpha|^2e^{-i|k|(t-t')}+|\beta|^2 e^{i|k|(t-t')}
+\alpha\beta e^{-i|k|(t+t')}+\alpha^*\beta^*e^{i|k|(t+t')}\Big), \label{eq:xpropag}
\end{align}
where we have written down the $SO(6)$ and the color indices explicitly. In what follows we will again suppress these indices. In the rest of this paper we consider a large quench limit (as was used in \cite{Sotiriadis:2010si}), where $m_0\gg1/L$, where $L$ is a characteristic length scale that we are interested in. More precisely, when $|x|-(t-t')\gg 1/m_0$, the propagator can be simplified
\beq
\langle\Phi(t,x)\Phi(t',0)\rangle\approx g^2\int_0^{\infty}\frac{dk}{8\pi^2|x|}\sin(k|x|)\frac{m_0}{k}\Big(\cos k(t-t')-\cos k(t+t')\Big)
.\nonumber
\eeq
By performing the integral using the identities in Appendix \ref{sec:integrals} we obtain
\beq
K(t,x;t',0)=\langle\Phi(t,x)\Phi(t',0)\rangle\approx\frac{g^2m_0}{16\pi}\frac{1}{|x|}\Theta(|x|,t,t'),\label{eq:wightman}
\eeq
where we have defined
\beq
\Theta(|x|,t,t')=\frac{1}{2}(\textrm{sgn}(|x|+t-t')+\textrm{sgn}(|x|-t+t')-\textrm{sgn}(|x|+t+t')-\textrm{sgn}(|x|-t-t')).
\eeq
This function is visualized in Figure \ref{fig:step}. The two point function (\ref{eq:wightman}) is the Wightman function, but it should be noted that in the deep quench approximation, the expectation value of the field commutator is subleading in $1/m_0$, making the Wightman functions and time ordered two point function equivalent, in the leading order.

The approximation leading to (\ref{eq:wightman}) breaks down near the light cone $|x|-(t-t')=0$. Instead, the two point function has the same singularities on the light cone
as in the vacuum state. These singularities are likely irrelevant for the qualitative behavior of the Wilson loop. We do not have methods to deal with the singular parts quantitatively. Thus, we will ignore them for most of the calculations and comment on their effects in Appendix \ref{sec:lightcone}.

\section{Climbing the first ladder}

\subsection{Specifying the Wilson loop}\label{sec:wilson}

\begin{figure}[h]
\begin{center}
\includegraphics[scale=.7]{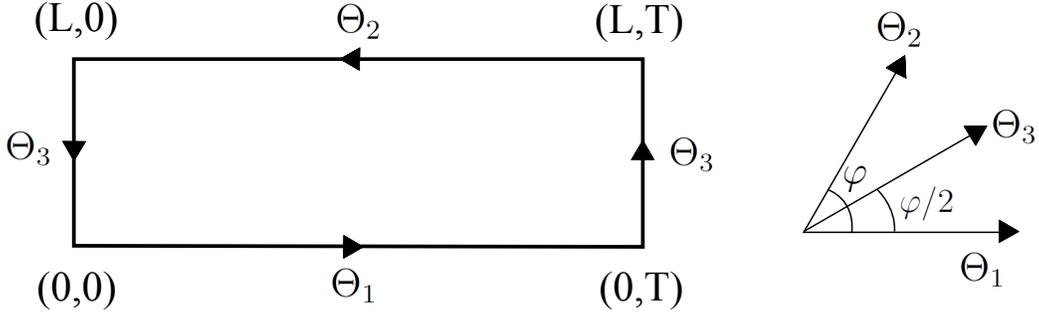}
\caption{\label{fig:loop} The rectangular Wilson loop, with $SO(6)$ orientations denoted as $\Theta$'s.}
\end{center}
\end{figure}

To specify a real time Wilson loop one needs to specify a curve in spacetime. In the current paper we work in Minkowski spacetime $R^{3,1}$.
For most of this paper we will choose to study rectangular Wilson loops whose sides have a temporal length $T$ and a spatial length $L$, shown in Figure \ref{fig:loop}. In this paper we consider the Maldacena-Wilson loop \cite{Maldacena:1998im}
\beq
\langle W(C) \rangle =\frac{1}{N}\langle\textrm{Tr}Pe^{i\int_C d\tau( A_{\mu}\dot{x}^{\mu}-\Phi_i\theta_i\sqrt{-\dot{x}^2})}\rangle,\label{eq:wilson}
\eeq
where $P$ denotes path ordering of the matrices. This is the Wilson loop that sources a fundamental string in the dual string theory \cite{Maldacena:1998im}. To define the Wilson loop
(\ref{eq:wilson}) one has to specify a direction of the symmetry breaking VEV of the (external) W-boson in the R-symmetry group $SO(6)$.
We will choose the relative angle between the upper timelike segment and the lower timelike segment to be $\varphi$. While the
relative angle between the spacelike segments and the lower segment is chosen to be $\varphi/2$. The choices of the $SO(6)$ orientations
are displayed in Figure \ref{fig:loop}.

In what follows we use the approximation introduced in \cite{Correa:2012nk} which isolates the scalar ladder diagrams as the leading contribution
to the Wilson loop. For $n$ $\Phi$ propagators going between the timelike segments we have a factor of $(\cos(\varphi))^n$. For the planar ladder
diagrams there is also a corresponding factor of $\lambda^n$. More general diagrams with more interactions on the internal lines
have more powers of $\lambda$. This suggests a limit which picks up the scalar ladder diagrams
\beq
\varphi=i\vartheta,\quad \vartheta\rightarrow\infty,
\eeq
while keeping
\beq
\hat{\lambda}=\frac{1}{4}e^{\vartheta}\lambda,
\eeq
fixed. Diagrams with lines running between spacelike and timelike lines are suppressed in this limit as $\hat{\lambda}e^{-\vartheta/2}$,
while the lines running between the spacelike segments are suppressed as $\hat{\lambda}e^{-\vartheta}$.
All the corrections to the free scalar propagator are suppressed by powers of $\hat{\lambda}e^{-\vartheta}$. It is likely that the
approximation of neglecting such diagrams will not commute with the large distance limit.

There is some freedom in defining a real time Wilson loop. One can consider several different orderings of the operators (see e.g. \cite{Tseytlin:2002tr,Laine:2006ns}). One could for example consider a time ordered operator ordering, or a path ordered one with some choice for
the starting point of the path. In the leading order deep quench approximation, the time ordered two point function is equivalent to the Wightman functions, making these choices to give the same result for the Wilson loop. Thus, we can simply ignore this subtlety in this work.

\subsection{1-loop warmup}

First consider the first order correction to the Wilson loop as a warmup. This is given by
\beq
\langle W(C)\rangle\approx1-\frac{\theta_i\theta_j\textrm{Tr}(T^{a}T^{b})}{N}\int_0^Tdt_1\int_0^T dt_2\sqrt{-\dot{x}_1^2}\sqrt{-\dot{x}_2^2}
\langle \Phi^a_i(x_1)\Phi^b_j(x_2)\rangle.
\eeq
Using $T^aT^a\approx N/2$ and the two point function (\ref{eq:wightman}) we get
\beq
\langle W(C)\rangle\approx 1-\frac{\hat{\lambda}}{16\pi }\int_0^Tdt\int_0^S ds \frac{m_0}{L}\Theta(L,t,s)
.
\eeq
Rather than giving the full result for the integral we note the most important things about the answer.
Firstly the whole correction vanishes for $2T-L<0$, since $\Theta(L,t,s)$ vanishes within the corresponding integration region.
Secondly, for large times $T\gg L/2$ we find
\beq
\langle W(C)\rangle\approx 1-\frac{m_0\hat{\lambda}}{8\pi}T.\label{eq:1loop}
\eeq
Since the two point Wightman function (\ref{eq:wightman}) is real, we immediately see that the Wilson loop
will be real. We will come back to this point later on.

\section{Summing up the ladders}

\subsection{The Bethe-Salpeter equation}

It is well known that ladder diagrams can be resummed using a Bethe-Salpeter equation \cite{Salpeter:1951sz,Erickson:1999qv,Erickson:2000af}.

\begin{figure}[h]
\begin{center}
\includegraphics[scale=1]{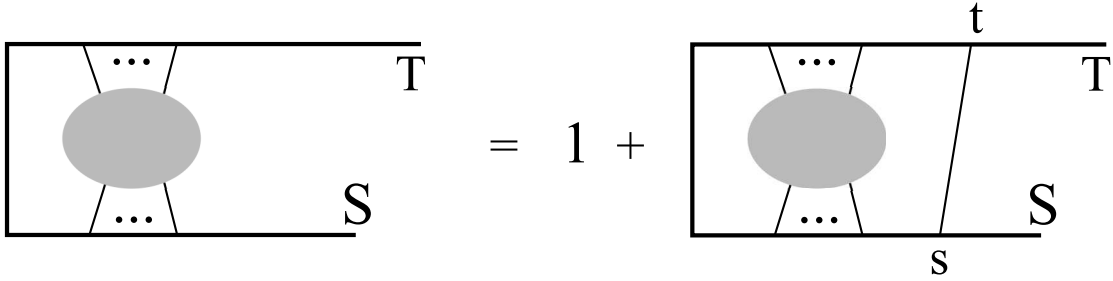}
\caption{\label{fig:Bethesalpeter} The Bethe-Salpeter equation resumming the ladder diagrams.}
\end{center}
\end{figure}

The Bethe-Salpeter equation in question is graphically shown
in Figure \ref{fig:Bethesalpeter}. As an equation it reads
\beq
\Gamma(S,T)=1-\frac{N}{2}\cos\varphi\int_0^Sds\int_0^Tdt K(t,L;s,0)\Gamma(s,t),\label{eq:bs1}
\eeq
where $K$ is the two point function (\ref{eq:wightman}).
Taking two derivatives of (\ref{eq:bs1}) gives a differential equation for $\Gamma$
\beq
\frac{\partial^2\Gamma(S,T)}{\partial S\partial T}=-\frac{N}{2}\cos\varphi K(T,L;S,0)\Gamma(S,T).
\eeq
This equation should be solved with the boundary conditions
\beq
\Gamma(S,0)=1,\quad \Gamma(0,T)=1,
\eeq
These boundary conditions follow from the integral form (\ref{eq:bs1}).

\begin{figure}[h]
\begin{center}
\includegraphics[scale=.7]{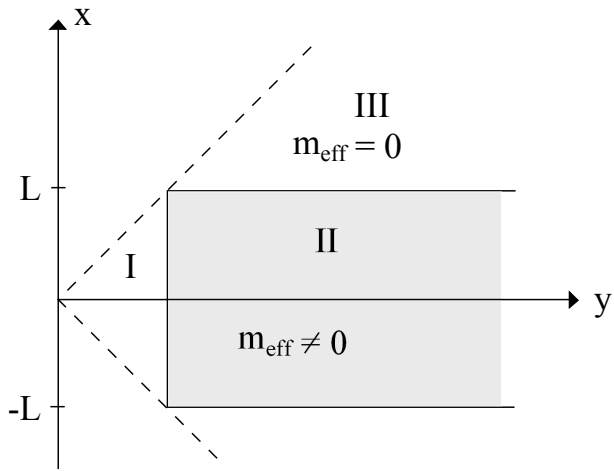}
\caption{\label{fig:waveeq} The Klein-Gordon equation (\ref{eq:waveeq}) is massless in the regions I and III. The effective mass
is non-vanishing in the region II. The dashed lines correspond to the light-cone where the initial conditions are specified.}
\end{center}
\end{figure}

A more familiar form of the equation is
obtained by defining $x=S-T$ and $y=S+T$. This way the equation becomes
\beq
(\partial_y^2-\partial_x^2+m_{eff}^2)\Gamma(x,y)=0\label{eq:waveeq},
\eeq
which is nothing but the Klein-Gordon equation with a spacetime dependent mass term
\beq
m_{eff}^2=\frac{\hat{\lambda}m_0}{16\pi L}\Theta\Big(L,\frac{y+x}{2},\frac{y-x}{2}\Big)\equiv \mu^2\Theta\Big(L,\frac{y+x}{2},\frac{y-x}{2}\Big).
\eeq
The form of the mass term is shown in Figure \ref{fig:waveeq}. The boundary condition for the Klein-Gordon equation is given at the light-cone $x=\pm y$ with $y\ge0$.

\subsection{The horizon effect}

First consider solving the BS equation (\ref{eq:waveeq}) in the region I shown in Figure \ref{fig:waveeq}. In this region, it reduces
to a massless Klein-Gordon equation.

The massless Klein-Gordon equation is most conveniently analysed using the original coordinates $S$ and $T$. The
equation reads
\beq
\partial_S\partial_T\Gamma=0.
\eeq
The general solution is simply
\beq
\Gamma=f(T)+g(S),
\eeq
where the functions $f$ and $g$ are fixed by the boundary conditions $\Gamma(S,0)=\Gamma(T,0)=1$, which gives
\beq
f(0)+g(T)=1,\quad f(S)+g(0)=1.
\eeq
Thus, $f$ and $g$ must be constants, that satisfy $f+g=1$ ,which fixes
\beq
\Gamma(S,T)=f+g=1.
\eeq
This immediately tells us that in the early time region $y<L$, where $m_{eff}=0$, the solution is simply $\Gamma=1$.

This means in particular that all of the ladder diagram corrections to the Wilson loop vanish when $T<L/2$. This is analogous to the "horizon effect" noticed for correlation functions in \cite{Calabrese:2007rg} and for entanglement entropy in \cite{Calabrese:2005in}. The "horizon effect" reflects the fact the in order for the quench to generate correlations between distant points, one at least has to wait for the time $T=L/2$ for
the correlations between spacelike separated operators to be causally affected by the quench.


\subsection{A Heuristic picture of the decay of the Wilson loop}\label{sec:heuristic}

To understand the qualitative behavior of the solutions of the Klein-Gordon equation (\ref{eq:waveeq}) for later times, it is convenient to introduce a $1+1$ dimensional energy momentum tensor for the field $\Gamma$
\beq
T_{\mu\nu}=-\frac{1}{2}\eta_{\mu\nu}(\partial\Gamma)^2+\partial_{\mu}\Gamma\partial_{\nu}\Gamma-\frac{1}{2}\eta_{\mu\nu}m_{eff}^2\Gamma^2.
\eeq
The four momentum density $T_{\mu 0}$ is conserved separately for times $y<L$ and $y>L$, while its value jumps at $y=L$.
For $y<L$ the solution that satisfies the appropriate boundary conditions is $\Gamma=1$. Thus, the energy of the fields vanishes for $y<L$.
As the time passes $y=L$, the mass term gets a non-vanishing value and the system obtains an energy
\beq
E=\int_{-L}^{L}T_{00}=\mu^2L.
\eeq
This will make the field oscillate on top of the potential hill in the region $-L<x<L$ where the mass is non-vanishing. Ignoring the regions $x>L$ and $x<-L$, this oscillation would not be decaying but would have the form $\Gamma= \cos(\mu(y-L))$. Taking into account the fact that the field can leak out of the
region $-L < x < L$ leads to a decay of the solution. To get a more quantitative feel of how the field behaves we make an ansatz
that the field inside the massive region has the form
\beq
\Gamma\approx A(y)\cos(\mu(y-L)),
\eeq
where $A(y)$ is assumed to be a slower varying function than the cosine. Taking into account the fact that the energy in the field oscillations can escape from the massive region we obtain an outgoing energy flux
\beq
T_{yx}=\partial_x\Gamma\partial_y\Gamma\propto \mu^2 A^2,
\eeq
where we used the fact that the waves outside $-L<x<L$ satisfy a massless wave equation so that $\partial_x\Gamma=\pm \partial_y\Gamma$ at $x=\pm L$. This decreases the total energy inside the massive region $E_L\propto \mu^2 A^2 L$ with a rate
\beq
\frac{\partial E_L}{\partial y}\propto -\mu^2 A^2,
\eeq
which gives a differential equation for $A(y)$ that is solved by
\beq
A(y)\propto e^{-c\frac{y}{2L}}.
\eeq
This leads to the following form for the Wilson loop
\beq
\langle W(C)\rangle\propto \cos(2\mu T)e^{-c T/L}.
\eeq
The coefficient $c$ is a constant factor (meaning independent of $T$) that is not determined by our heuristic calculation. Thus, we see that the qualitative behavior of the Klein-Gordon field is the following. There are oscillations in the field in the region $-L < x < L$ that escape through the boundaries $x=\pm L$ leading in to a loss of energy and an exponential decay of $\Gamma$. This behavior is characteristic to a system exhibiting quasinormal modes. In the next section we make this connection more precise.

At first one might be surprised that the real time Wilson loop does not have the usual $e^{-iTE}$ behavior, but rather behaves as $e^{-T\gamma}$.
Such an exponential decay is often a sign of an instability \cite{Tseytlin:2002tr,Laine:2006ns}. In the current case it could be interpreted as signalling that mesons become unstable in an energetic bath of particles as was found in thermal equilibrium in pure Yang-Mills theory in \cite{Laine:2006ns}.

\subsection{Reducing the Wilson loop to a quasinormal mode problem}\label{sec:quasinormal}

\begin{figure}[h]
\begin{center}
\includegraphics[scale=1]{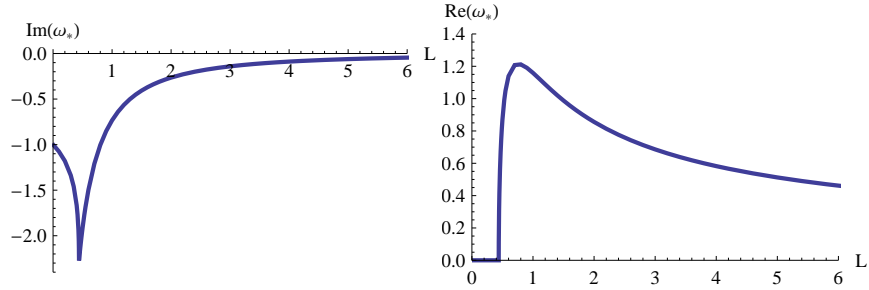}
\caption{\label{fig:eigenvalue} Solutions to the eigenvalue equation obtained numerically as functions of $L$. Left: The imaginary part of the lowest quasinormal frequency. Right: The real part of the lowest quasinormal frequency. The units in the figure are chosen so that $\hat{\lambda} m_0/16\pi=1$.}
\end{center}
\end{figure}

To obtain a fully quantitative solution of (\ref{eq:waveeq}) it is convenient to solve the fields outside the potential hill $-L<x<L$.
This can be done by noting that the the solution in the region $x>L$ is
\beq
\Gamma=\Gamma_+=g_+(y-x)+1-g_+(0),
\eeq
and correspondingly in $x<-L$
\beq
\Gamma=\Gamma_-=f_-(y+x)+1-f_-(0).
\eeq
The only data in these regions is the amount of outgoing radiation going towards $x=\pm\infty$. Recalling that the Wilson loop is obtained as
the value of $\Gamma$ at $x=0$, suggests we do not need to know the form of the outgoing radiation and can concentrate on the region $-L<x<L$. This can be done by noting that at $x=\pm L$, $\Gamma$ satisfies
\beq
(\partial_x\pm\partial_y)\Gamma=0.\label{eq:boundary1}
\eeq
When $y>L$, the effective mass term is independent of $y$, which allows us to perform a Fourier transform
\beq
\Gamma(y,x)=\int d\omega e^{-iy\omega}\tilde{\Gamma}(\omega,x).
\eeq
This way the field equation (\ref{eq:waveeq}) and the condition (\ref{eq:boundary1}) can be reduced into a massive wave equation within
the region $x\in(0,L)$ together with a mixed boundary condition
\beq
(-\partial_x^2-\omega^2+\mu^2)\tilde{\Gamma}=0,\quad \partial_x\tilde{\Gamma}(x=0)=0,\quad \partial_x\tilde{\Gamma}(x=L)= i\omega\tilde{\Gamma}(x=L),\label{eq:quasi1}
\eeq
where we used the symmetry of the problem under $x\rightarrow -x$. Of course equations (\ref{eq:quasi1}) are accompanied with the initial data that $\Gamma(x,y=L)=1$. We can look for mode solutions to the equation (\ref{eq:quasi1}) which have the form
\beq
\tilde{\Gamma}=A\cos(\sqrt{\omega^2-\mu^2}x).
\eeq
Imposing the boundary condition at the leaking wall $x=L$ leads to a condition for the frequency $\omega$ given by
\beq
\tan(\sqrt{\omega^2-\mu^2}L)=-i\frac{\omega}{\sqrt{\omega^2-\mu^2}}.\label{eq:eigenvalue}
\eeq
There are no real eigenvalues satisfying (\ref{eq:eigenvalue}) and we have been unable to find solutions on the upper half complex $\omega$ plane, as expected on physical grounds.
Thus, we indeed find that we are dealing with a quasinormal mode problem. Generically solutions to (\ref{eq:eigenvalue}) can be found numerically.
The real and imaginary parts of the quasinormal frequencies are shown in Figure \ref{fig:eigenvalue}.
The quasinormal modes exhibit a "phase transition" at $L=L_c\approx 22/\hat{\lambda}m_0$. When $L<L_c$, the modes are purely imaginary. While for $L>L_c$, the quasinormal frequency obtains a real part leading to an oscillation of the Wilson loop. This follows from the eigenvalue equation (\ref{eq:eigenvalue}), as two of the lowest quasinormal frequencies on the imaginary axis get closer to each other as $L$ is increased, and at $L=L_c$ they collide and move off the imaginary axis.

The lowest quasinormal mode at large $L$ can be found analytically as
\beq
\textrm{Re}(\omega_*)\approx\mu,\quad \textrm{Im}(\omega_*)\approx-\frac{4\pi^3}{\hat{\lambda}m_0}\frac{1}{L^2}.
\label{eq:largeL}
\eeq
So at large $L$ we find that the Wilson loop behaves as
\beq
\langle W(C)\rangle\approx\cos(2\mu T)e^{-T\frac{8\pi^3}{\hat{\lambda}m_0}\frac{1}{L^2}}.
\eeq
For small $L$ one can solve the Klein-Gordon equation (\ref{eq:waveeq}) exactly as described in Appendix \ref{sec:short}, which leads to
\beq
\langle W(C)\rangle=e^{-\frac{\hat{\lambda}m_0}{8\pi}\theta(T-\frac{L}{2})(T-\frac{L}{2})},
\eeq
in agreement with the numerical solution of the eigenvalue equation shown in Figure \ref{fig:eigenvalue} and also with the
1-loop result (\ref{eq:1loop}).

\subsection{Numerical solutions}

One can also find full numerical solutions of the wave equation (\ref{eq:waveeq}). This way we can check that the approximate solutions discussed in the previous sections indeed corresponds to the full real time evolution.

The numerical solutions are obtained using Mathematica's NDSolve. For $L\mu$ of order 1 we indeed find good agreement with the numerical solutions and the previous approximate solutions.\footnote{For $L \mu \gg 1$ our numerical methods of solving (\ref{eq:waveeq}) become unreliable.} In particular we find that the exponential decay of $\Gamma$ starts quickly after the time $y=L$. Heuristically this is indeed expected as the walls of the system leak right from the beginning, leading to inevitable exponential decay.

\begin{figure}[h]
\begin{center}
\includegraphics[scale=1]{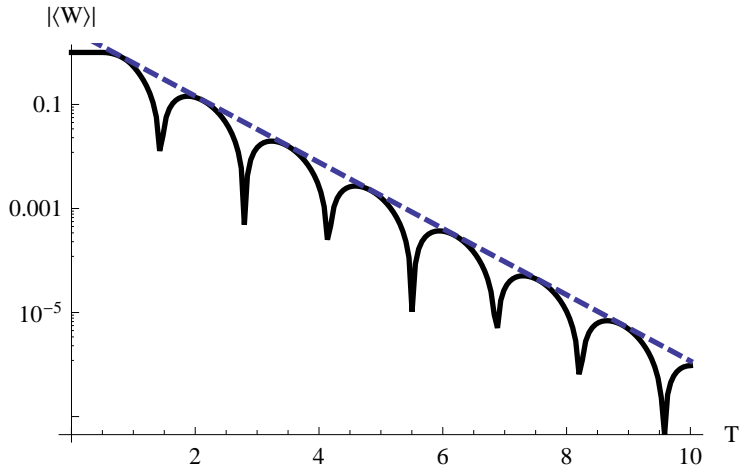}
\caption{\label{fig:example} An example numerical solution of (\ref{eq:waveeq}) shown in log scale. The dashed line corresponds to the imaginary part of the quasinormal mode obtained as a solution to (\ref{eq:eigenvalue}). The figure corresponds to the choice of parameters $L m_0=16\pi /\hat{\lambda}$}
\end{center}
\end{figure}

An example numerical solution together with the exponential decay predicted by solving the eigenvalue equation (\ref{eq:eigenvalue}) are shown in Figure \ref{fig:example}. The solution is seen to decay exponentially with a rate given by the imaginary part of the lowest quasinormal mode. Also the oscillation period is found to be given by the real part of the lowest quasinormal mode to a good accuracy.

\subsection{Comparison to thermal equilibrium}\label{sec:thermal}

In this section we compare the non-equilibrium Wilson loops to real time thermal Wilson loops.\footnote{Euclidean time thermal Wilson loops have been studied in the ladder approximation in \cite{Shuryak:2003ja}, with and without thermal screening.} We will continue to work in the large $\vartheta$ limit and also take a high-temperature limit. The finite temperature real time Wightman function with temperature $\beta^{-1}$ is given by
\beq
K(t,x;t',0)\approx\frac{g^2}{4\beta\pi|x|}\Theta_{\beta}(x,t),\label{eq:2pt}
\eeq
where we have ignored terms that grow slower than $1/\beta$, as $\beta\rightarrow 0$, as we are interested in comparing to the large quench results which hold in a limit of a large energy density. Above we have defined
\beq
\Theta_{\beta}(x,t)=\frac{1}{2}(\textrm{sgn}(|x|+t-t')+\textrm{sgn}(|x|-t+t')).
\eeq
As the large temperature two point function (\ref{eq:2pt}) has a very similar form as the non-equilibrium two point function, the subsequent steps leading to a Bethe-Salpeter equation are identical as before. We are lead to a Klein-Gordon like equation for the finite temperature Wilson loop $\Gamma_{\beta}$
\beq
(\partial_y^2-\partial_x^2+m_{\beta}^2)\Gamma_{\beta}=0,
\eeq
where now
\beq
m_{\beta}^2=\frac{\hat{\lambda}}{4\pi\beta L}\Theta_{\beta}(L,x).
\eeq
Furthermore, all our results on exponential decay of the Wilson loops apply to the thermal case, with the only difference being the initial conditions, which basically just means that the "horizon effect" is absent from the thermal state as the thermal state is time translationally invariant.

Thus, the non-equilibrium Wilson loops decay with the same quasinormal spectrum as in the thermal case if we identify the temperature $1/\beta=m_0/4$, which means that they are practically thermal once the time has passed $T=L/2$.

\section{Spacelike Wilson loops}

In this section we consider spacelike rectangular Wilson loops, with sides $x_1\in (0,L)$ and $x_2\in (0,Z)$. We choose the $SO(6)$ orientations as described in section
\ref{sec:wilson}, with the replacement $T\rightarrow Z$, so that only the ladder diagrams between the constant $x_1$ lines contribute. In this case there are no ambiguities with real time
time ordering as the loop is located at constant time. The two point functions that contribute to the ladders are equal time two point functions. The Bethe-Salpeter equation in this case becomes
\beq
\Gamma(Z_1,Z_2)=1+\frac{\hat{\lambda}m_0}{16\pi}\theta(t-L/2)\int_0^{Z_1}dz_1\int_0^{Z_2}dz_2\frac{\Gamma(z_1,z_2)}{\sqrt{L^2+(z_1-z_2)^2}}.\label{eq:spacelikeBS}
\eeq
To find out the time dependence of the spacelike Wilson loop, we do not even have to solve (\ref{eq:spacelikeBS}). We can simply note that
for times $t<L/2$ the integral term vanishes and we obtain $\Gamma=1$. This again gives us the horizon effect for spacelike Wilson loops.
For later times $t>L/2$, the Bethe-Salpeter equation (\ref{eq:spacelikeBS}) is independent of time and thus $\Gamma$ stays constant. Thus, as far as the time dependence is concerned the Wilson loop has the form
\beq
\langle W\rangle=e^{\theta(t-L/2)f(L,Z)}.
\eeq
The $L$ dependence can be worked out for example by using the methods of \cite{Erickson:1999qv}. The function $f(L,Z)$ is again the same as one would obtain at finite temperature with the identification $\beta^{-1}=m_0/4$.

By taking two derivatives of (\ref{eq:spacelikeBS}) we obtain the differential equation
\beq
\partial_{Z_1}\partial_{Z_2}\Gamma(Z_1,Z_2)=\gamma\frac{\Gamma(Z_1,Z_2)}{\sqrt{L^2+(Z_1-Z_2)^2}},
\eeq
where we denote $\gamma=\hat{\lambda}m_0/(16\pi)$ and are considering the case $t>L/2$. Again we can use a change of coordinates $x=Z_1-Z_2$ and $y=Z_1+Z_2$ to rewrite the equation as
\beq
(\partial_y^2-\partial_x^2)\Gamma(x,y)=\gamma\frac{\Gamma(x,y)}{\sqrt{L^2+x^2}}.\label{eq:spatial2}
\eeq
This equation can be straightforwardly solved using numerical methods. Here we consider the limit of large $\gamma$, in
which case an analytic solution can be obtained. In this limit, the potential well on the right hand side becomes very
deep and can be approximated by a quadratic potential as in \cite{Erickson:1999qv}. Furthermore to obtain the
large $Z$ behavior it is sufficient to find the Laplace transformed solution $\Gamma=e^{\omega y}\psi(x)$ with
the largest eigenvalue $\omega$. Thus, we are lead to solve the ground state energy of the harmonic oscillator
\beq
(-\partial_x^2+\frac{\gamma}{2L^3}x^2)\psi(x)=(\frac{\gamma}{L}-\omega^2)\psi(x).
\eeq
The ground state wavefunction of the oscillator is simply
\beq
\psi\propto e^{-\sqrt{\frac{\gamma}{8}}\frac{x^2}{L^{3/2}}},
\eeq
and the minimum eigenvalue is
\beq
\omega=\sqrt{\frac{\gamma}{L}}-\frac{1}{2\sqrt{2}L}+\mathcal{O}(1/\sqrt{\gamma}).
\eeq
Thus we obtain
\beq
\langle W\rangle \propto \exp\Big[Z\Big(\sqrt{\frac{\hat{\lambda}}{\pi\beta L}}-\frac{1}{\sqrt{2}L}\Big)\Big],\label{eq:spatialwilson}
\eeq
where $\beta=4/m_0$ is the effective temperature. The result (\ref{eq:spatialwilson}) is quite different from strong coupling result for the finite temperature spacelike
Wilson loop obtained from a string theory calculation on a black hole background. The strong coupling result has the form \cite{Witten:1998zw,Brandhuber:1998er}
\beq
\langle W\rangle \propto e^{-c \sqrt{\lambda}\beta^{-2} Z L},\label{eq:string}
\eeq
with an order one coefficient $c$. The area law reflects the fact that the dimensionally reduced thermal field theory,
when interpreted as a 3 dimensional Euclidean quantum field theory,
confines at strong coupling \cite{Witten:1998zw,Brandhuber:1998er}.
Thus, it is not surprising that the $L$ dependence from weak coupling is different from the strong
coupling result.
Imaginary separation of the string on the boundary $S^5$ does not change this result in the limit of large temperature, as can
be checked by a direct calculation.\footnote{The string calculation suffers from a potential order of limits problem.
The other order of limits, which is to first take the imaginary separation large and then take the temperature large,
can lead to a result different from (\ref{eq:string}). We leave the full study of this problem to future work. The first order of
limits is the one we use at weak coupling, as we first used the deep quench approximation before taking $\hat{\lambda}$ to be large.}

\newpage

\section{Discussion}

\begin{figure}[h]
\begin{center}
\includegraphics[scale=.7]{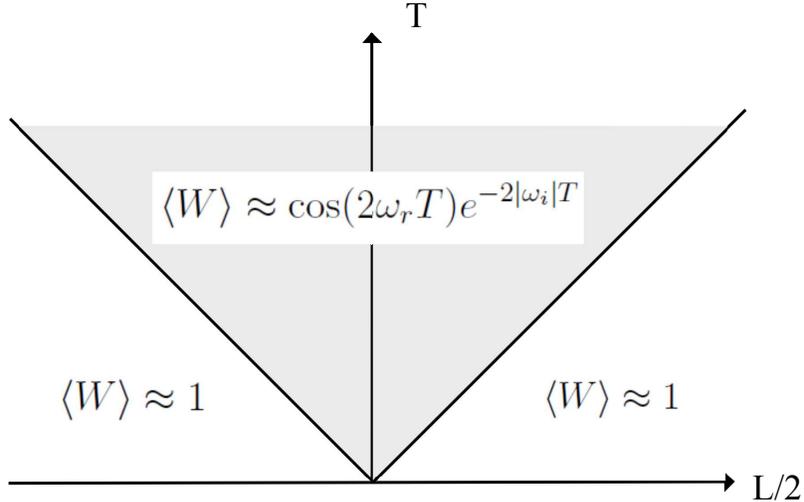}
\caption{\label{fig:result} The qualitative form of the results for the timelike Wilson loop. Here $\omega_*=\omega_r+i\omega_i$ are the quasinormal frequencies, which are functions of $L$.}
\end{center}
\end{figure}

The results of our calculations are the following. The Wilson loops exhibit a "horizon effect" similar to that found for two point functions and entanglement entropy in 1+1 conformal field theories \cite{Calabrese:2005in,Sotiriadis:2010si}. More precisely this means that the Wilson loop is simply 1 in the region $T<L/2$. The "horizon effect" reflects the fact the in order for the quench to generate correlations between distant points, one at least has to wait for the time $T=L/2$ for
the correlations between spacelike separated operators to be causally affected by the quench.

Furthermore we find that the timelike Wilson loop decays with quasinormal modes identical to those in thermal equilibrium, with the temperature identified as $\beta^{-1}=m_0/4$.
The spacelike Wilson loop is found to reach a thermal value immediately after $t>L/2$. The result for the timelike Wilson loop is summarized in Fig. \ref{fig:result}.

Thermalization of the Wilson loops may seem surprising since the calculations we have performed involve really only free field theory. The mathematical reason
for it follows from the fact that the late time scalar two point correlation function approaches the form of a thermal two point function with the identification of the
effective temperature $\beta^{-1}=m_0/4$. This happens for correlators with spatial separation $L$ satisfying $2T\gg L\gg\beta$. For larger length scales the
correlation function is far from equilibrium. This is simply the "horizon effect" \cite{Calabrese:2005in,Sotiriadis:2010si}. For length scales shorter than $\beta$,
the two point function in the quench state is non-thermal. Thus, it is likely that the Wilson loops with $L<\beta$ are also non-thermal in the quench state. Unfortunately
this region is beyond the deep quench approximation we have used in this work.

When interactions are taken into account, by including Feynman diagrams that are not of the ladder form, the free thermal form of the Wilson loops becomes only a temporary stage of the time evolution and final thermalization is achieved in a parametrically larger timescale due to the interactions. The form of the two point function is known to change qualitatively at large times and distances as interactions are taken into account. In our case the one loop corrections are expected to generate a distance/time scale of the order $1/m_0\sqrt{\lambda}$, which can be assumed to be large at weak coupling. Thus, our calculation only describes the time evolution of the Wilson loops in time scales small compared to $1/m_0\sqrt{\lambda}$. At the time $T\approx L/2$ after the quench, the timelike Wilson loops start decaying with thermal quasinormal modes, and the spacelike Wilson loop relaxes into a static thermal form. The thermal character of the expectation values follows from the choice of the initial state. Then, as time becomes of the order of $1/m_0\sqrt{\lambda}$, interactions will start changing the picture and one might expect that gradually the system will fully thermalize into the thermal state of the interacting theory due to interactions between particles. Our calculation clearly only applies to the early time behavior of the quench dynamics of the interacting theory.

In the following we will briefly compare the qualitative results found at weak coupling, with the corresponding strong coupling picture found from black hole physics.
In the strong coupling limit the physics of the thermalization can be studied through semiclassical supergravity in the bulk. In the bulk picture fast varying time dependent sources are expected to lead to the creation of a shell of energy density
near the boundary, which subsequently falls into $AdS$ forming a black hole \cite{Bhattacharyya:2009uu,Wu:2012rib}.

Correlation functions\cite{Balasubramanian:2010ce,Balasubramanian:2011ur}, entanglement entropy \cite{AbajoArrastia:2010yt} and Wilson loops \cite{Balasubramanian:2010ce,Balasubramanian:2011ur} have been studied in such falling shell backgrounds.
Perhaps the main qualitative lessons arising from the collapsing shell calculations is that correlations separated by a spacelike distance $L$ thermalize maximally fast in time $T\approx L/2$ \cite{Balasubramanian:2010ce,Balasubramanian:2011ur}, with a memory on the initial state being of the order of the time scale of the lowest quasinormal mode \cite{Chesler:2011ds,Bhaseen:2012gg}.

At strong coupling \cite{Balasubramanian:2010ce,Balasubramanian:2011ur}, short spacelike Wilson loops thermalize faster than long Wilson loops, with the thermalization time approximately satisfying $T\approx L/2$. This is easily understood, as a long string can pass through the falling shell surface seeing the time dependent geometry for longer times. Thus, our weak coupling calculation reproduces this basic pattern.

The $L$ dependence of the spatial Wilson loop is very different at weak and strong couplings. The thermal strong coupling result shows an area
law $\log\langle W\rangle\propto -ZL$, signalling magnetic confinement \cite{Witten:1998zw,Brandhuber:1998er}. This is certainly not the case at weak coupling,
where we have found $\log\langle W\rangle\propto Z/\sqrt{L}$. The disagreement between the $L$ dependence of the weak and strong coupling results is not
particularly surprising.

The exponential decay of the timelike Wilson loop that we have found might seem surprising at first as compared to the usual functional form $e^{-iET}$. Such exponential decay has been noticed for Wilson loops for example in \cite{Tseytlin:2002tr,Laine:2006ns}. It seems likely that in our case it is signalling an instability of heavy mesons in the energetic bath of particles produced by the quench.

From the strong coupling point of view, it seems plausible that real time Wilson loops indeed decay exponentially in time in a black hole background. It is well know that particle propagators decay exponentially (with quasinormal behavior) in black hole backgrounds as propability falls through the horizon.
The Wilson loop may be thought of as an amplitude for a flat string lying at $z=\epsilon$ and $t=0$ extended in $x$ direction with a length $L$ to be found at the same position at a later time $t=T$, keeping the endpoints fixed at the boundary. As long as the string is sufficiently long one would expect to find a similar leaking of propability through the horizon, leading to exponential decay. Results supporting the exponential decay have been indeed recently found in \cite{Hayata:2012rw,Finazzo:2013rqy}.

Anther interesting question is whether a strong coupling calculation leads to the higher quasinormal modes controlling the decay of the Wilson loop. Due to the lack
of calculational methods, we will leave the study of this question for future work.

\section*{Acknowledgements}

I would like to thank J. Alanen, K. Kajantie, E. Keski-Vakkuri, K. Rummukainen, S. Stricker, O. Taanila, L. Thorlacius, A. Vuorinen and K. Zarembo for discussions. This work was supported in part by the Icelandic Research Fund and by the University of Iceland Research Fund.

\appendix

\section{Some useful integrals}\label{sec:integrals}

To perform Fourier transforms
\beq
f_n(\alpha)=\int_{-\infty}^{\infty}\theta(k)k^{-n}e^{ik\alpha},
\eeq
we use the following representation of theta function
\beq
\theta(k)=\int_{-\infty}^{\infty}\frac{ds}{2\pi i}\frac{e^{isk}}{s-i\epsilon}.
\eeq
This allows us to write
\beq
f_n(\alpha)=\int_{-\infty}^{\infty}\frac{ds}{2\pi i}\frac{1}{s-i\epsilon}\int_{-\infty}^{\infty}dk k^{-n}e^{i k(\alpha+s)}.
\eeq
Thus we have
\beq
\partial_{\alpha}^nf_n(\alpha)=\frac{i^{n+1}}{\alpha+i\epsilon}.
\eeq
We will need $f_n$ for $n=1$ in which cases we get
\beq
f_1(\alpha)=-\log(\alpha+i\epsilon)+\textrm{const.}
\eeq
A further useful identity is
\beq
f_1(\alpha)-f_1(-\alpha)=\log(-\alpha+i\epsilon)-\log(\alpha+i\epsilon)=i\pi \textrm{sgn}(\alpha).
\eeq

\section{Short distance approximation}\label{sec:short}

In the limit $L\rightarrow 0$ we can approximate the effective mas term with a delta function
\beq
m_{eff}^2=\frac{\alpha}{2L}\Theta(L,x,y)\rightarrow \alpha\theta(y-L)\delta(x),
\eeq
where we have introduced the mass scale $\alpha=\hat{\lambda}m_0/8\pi$. Denoting the region $x>0$ by + and $x<0$ by -, we can find a general solution to the massless Klein-Gordon equation
\beq
\Gamma=f_{\pm}(y+x)+g_{\pm}(y-x).
\eeq
Imposing the boundary condition $\Gamma(y,y)=1$ for $y>L$ and $\Gamma(y,-y)=1$ for $y>L$ leads to the solution
\beq
\Gamma=\Gamma_+=g_+(y-x)+1-g(0),
\eeq
on the region $+$ and
\beq
\Gamma=\Gamma_-=f_-(y+x)+1-f_-(0),
\eeq
on the region $-$. Requiring continuity at $x=0$ leads to
\beq
g_+(y)=f_-(y)+\textrm{const}.
\eeq
The equation of motion implies that the first $x$ derivative of $\Gamma$ has a discontinuity at $x=0$ given by
\beq
\partial_x\Gamma(y,x=-\epsilon)-\partial_x\Gamma(y,x=\epsilon)=-\alpha\Gamma(x=0,y).
\eeq
Using $\partial_x\Gamma(y,x=-\epsilon)=\partial_yf_-(y)$ and $\partial_x\Gamma(y,x=\epsilon)=-\partial_y f_-(y)$ gives a differential
equation for $f_-(y)$
\beq
\partial_y f_-(y)=-\frac{\alpha}{2}f_-(y),
\eeq
which is solved by
\beq
f_-(y)=f_0e^{-\alpha y/2}.
\eeq
Imposing the initial conditions leads to
\beq
\Gamma=e^{-\frac{\alpha}{2}\theta(y-L)(y-L)}.
\eeq
Thus, the Wilson loop in the large $L$ limit is given by
\beq
\langle W(C)\rangle=\Gamma(y=T,x=0)=e^{-\frac{\hat{\lambda}m_0}{8\pi}\theta(T-\frac{L}{2})(T-\frac{L}{2})}.
\eeq
Expanding the exponential in powers of $T$ we find that the first order indeed reproduces the perturbative result (\ref{eq:1loop}).

\section{Comments on the light-cone singularities}\label{sec:lightcone}

When calculating the Wightman function (\ref{eq:wightman}) we have taken the large $m_0$ limit in a rather naive way,
keeping only terms of the order $m_0$ in the two point function while neglecting the order $m_0^0$ contributions. At large distances/times
these contributions can be neglected as they provide only subleading contributions to the small $k$ singularities of the
momentum space two point function. There is still a situation where the order $m_0^0$ contributions (or even the order $m_0^{-1}$) cannot be
neglected. This is near the light cone singularities. There the two point function takes the same short distance form as in the ground state.
The contributions of the near lightcone region to the two point function have the form
\beq
\delta K\approx \frac{g^2}{|x|}c\delta_{m_0}(|x|-t),
\eeq
where $\delta_{m_0}$ is a linear combination of functions peaked around a region of the size $1/m_0$ and delta functions. These singularities
induce a correction term $\delta K$ to the Bethe-Salpeter kernel. The Klein-Gordon equation changes by the addition of walls of thickness $1/m_0$
to the edges of the potential hill at $x=\pm L$. Still, the waves can easily leak out of the potential hill leading to exponential decay of the Wilson loop as discussed in section \ref{sec:heuristic}.
To see more explicitly that the waves indeed pass through the walls we can consider a WKB approximation to the Klein-Gordon equation. Consider
an ansatz $\Gamma=e^{-i\omega t-i S(x)}$. In the WKB approximation, the tunneling through the walls is supressed through the imaginary part
\beq
|\textrm{Im}(S)|=\int_L^{L-x_0}dx\sqrt{m_{eff}^2(x)-\omega^2},\label{eq:WKB}
\eeq
where $L-x_0$ is the classical turning point with $x_0$ of the order $1/m_0$. The short distance form of $m_{eff}^2$ near the light cone is $m_{eff}^2\propto1/(x-L)$, which tells us that the WKB integral (\ref{eq:WKB}) is finite even though it passes through a singularity, and the tunneling is possible and not suppressed by any large negative exponent.

If the Klein-Gordon equation has bound states at the singular parts of $\delta K$, they can keep the value of the Wilson loop bounded from below by a factor of the order $e^{-c m_0 L}$, for some constant $c$. Studying these effects is outside the scope of this paper as they are highly suppressed in the deep quench limit.

\end{document}